\documentclass[prl,twocolumn,showpacs,amsmath,amssymb]{revtex4} \usepackage{color}
\usepackage{bm} \usepackage{graphics} \usepackage[pdftex]{graphicx} 

\begin{document} \frenchspacing

\title{\textbf{Quantum magnetoelectric effect in molecular crystal Dy$_3$}}

\author{D.I. Plokhov}
\affiliation{A.M. Prokhorov General Physics Institute of Russian Academy of Sciences, \\ 38 Vavilov Str., 119991, Moscow, Russia}

\author{A.I. Popov}
\affiliation{National Research University of Electronic Technology, \\ 5 Pas. 4806, 124498, Zelenograd, Moscow, Russia}

\author{A.K. Zvezdin}
\affiliation{A.M. Prokhorov General Physics Institute of Russian Academy of Sciences \\ 38 Vavilov Str., 119991, Moscow, Russia}

\date{\today}

\begin{abstract}
Magnetoelectric properties of a molecular crystal formed by dysprosium triangular clusters are investigated. The effective spin-electric Hamiltonian is derived on the base of developed quantum mechanical model of the cluster spin structure. The magnetoelectric contribution to the free energy of the crystal is calculated. The analysis reveals several distinctive features of the magnetoelectric effect, which are not typical for conventional paramagnetic systems at low temperatures. The peculiarities are explained by the chirality of the dysprosium core of the molecules.
\end{abstract}


\maketitle


\textbf{1.} Studying the properties of materials with new types of electronic structure, such as topological insulators \cite{Hasan}, or new types of charge or spin ordering, for example magnetic monopoles in spin ice \cite{Castelnovo}, is a tradition of condensed matter physics.

From this point of view, materials with toroidal magnetic ordering is of considerable interest \cite{Schmid,Spaldin,Kopaev}. The gender element of such ordering is the toroidal (anapole) moment, which is the main term in the toroidal family of current (or spin) system's multipole expansion \cite{Dubovik}. The simplest example of a system with toroidal moment is a toroidal solenoid. Toroidal moment ${\bf T}$ is a $t$-odd polar vector.

Several antiferromagnetic crystals with a toroidal type of magnetic ordering have already been found \cite{Popov,Sannikov,Krotov}. Such antiferromagnets reveal interesting magneto-optical and magnetoelectric properties. The presence of the toroidal moment provides a possibility of a magnetoelectric effect observation. According to the \cite{Spaldin,Kopaev}, induced polarization ${\bf P} \sim [{\bf H} \times {\bf T}]$.

Since recently, magnetoelectric and the related multiferroic materials have attracted considerable attention focused onto both improved fundamental understanding and novel desirable applications. Challenging and promising visions emerged, e.g. how to switch magnetism with bare electric fields and thus overcome the overheating bottleneck in microelectronic devices \cite{Lebeugle,Kleemann,Chen}.

In toroidal antiferromagnets, however, it is difficult to unambiguously separate effective quasi-isolated elements with toroidal moment the interaction between which would create toroidal moment of the substance. The toroidal moment in antiferromagnets is a collective property that is owed to the interspin exchange interaction.

Contrary to this, there is a molecular crystal consisting of Dy$_3$ metal-organic triangular molecular clusters \cite{Tang}, which are characterized by toroidal moment \cite{Soncini,EPL2009}. The cluster's distinctive feature is the zero magnetic moment of its ground state, despite the certain (clockwise or counterclockwise in the plane of the dysprosium triangle) arrangement of Dy$^{3+}$ ion spins. By the analogy with single molecular magnets (SMMs) we will call the Dy$_3$ clusters single molecular toroics (SMTs). Spin-electric effects in molecular antiferromagnets are recently reviewed in \cite{Trif}.

The presence of the intrinsic toroidal ordering in the Dy$_3$ system owes several peculiarities of the magnetoelectric properties which are qualitatively different from those of traditional materials. The investigation of the peculiarities is the purpose of this work.

First (see Sec. 2), we consider the spin structure of a Dy$_3$ molecule and show in what way the toroidal moment of the molecule is formed. The toroidal state appears to be degenerate. But (see Sec. 3) there is a possibility of preparing the non-generated states with the non-zero toroidal moment by the means of an applied external magnetic field.

Sec. 4 deals with the spin-electric interactions in the Dy$_3$ molecule. The spin-electric Hamiltonian is derived in the Appendix to this section and the magneto-electric corrections to the energy levels of Dy$_3$ molecule is obtained.

Sec. 5 presents the main results of our study. The magnetoelectric contribution to the free energy of the Dy$_3$ crystal is calculated and the magnetoelectric effect (MEE) is described in detail. The isotherms and field dependencies of the electric polarization are obtained for different ranges of magnetic field and temperature. The similarities and differences of the MEE in the system under study and conventional paramagnets (for example, rare-earth ferroborates) are discussed.

We should note that the whole Dy$_3$ crystal can be treated as a coherent array of the Dy$_3$ SMTs due to weak intermolecular interaction \cite{Chibotaru,Luzon}. This leads to the fact that macroscopic response of the crystal reveals the quantum properties of the very single molecule at low enough temperatures. Therefore, the magnetoelectric effect in Dy$_3$ molecular crystal bears traits of a macroscopic quantum coherent (MQC) effect.

Finally, Sec. 6 brings forward the unique possibility of observing a "purely" quantum non-equilibrium linear magnetoelectric effect in the Dy$_3$ molecular system making the old Pierre Curie's idea of MEE in molecules come true. \medskip


\textbf{2.} According to the experimental data \cite{Luzon,EPAPS}, a molecular crystal formed by triangular metalorganic nanoclusters Dy$_3$ \cite{Tang} is monoclinic with space group C2/c. One primitive cell contains eight Dy$_3$ molecules and is described by the parameters $a = 32.917(2)$ \AA, $b = 18.0213(13)$ \AA, $c = 17.3693(13)$ \AA, and $\beta = 114.522(1)^{\circ}$. The skeleton of each Dy$_3$ molecule is formed by three Dy$^{3+}$ ions, located in the apices of an equilateral triangle. The triangles are arranged in a regular way with the relative sides to be parallel. Similar to the SMMs, it is proved experimentally \cite{Tang,Luzon} that the Dy$_3$ molecules are very weakly coupled, thus the crystal can be considered as ''paramagnetic'', which means that the crystal is the ensemble of non-interacting dysprosium triangles with strong inter-ion exchange. Because the magnetoelectric properties of the crystal are owed to the symmetry of the triangle, we consider the three-ion system in a Dy$_3$ molecule.

\begin{figure} \centering \includegraphics[scale=0.35]{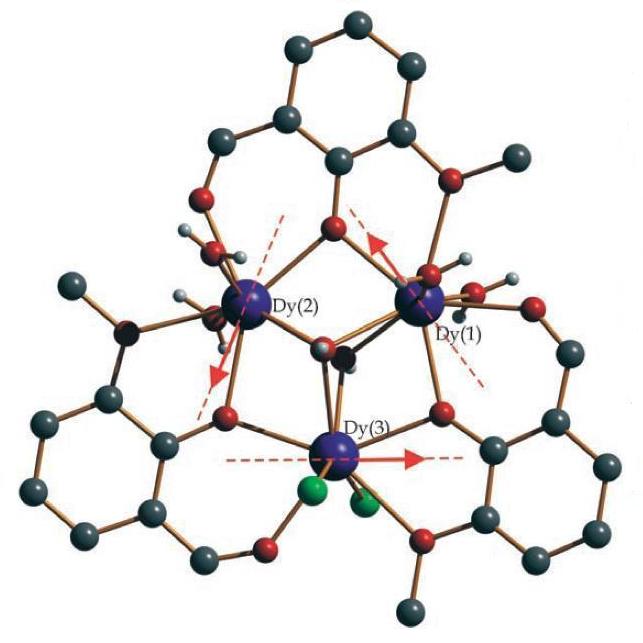} \includegraphics[scale=0.40]{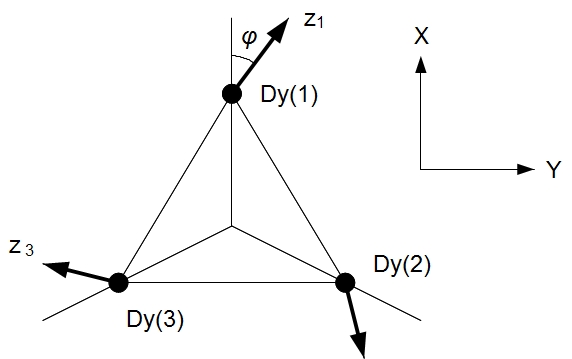} \caption{\label{dysp} (Upper) The structure of one of the three synthesized triangular Dy$_3$ nanoclusters. Color scheme: blue Dy, red O, green Cl, dark grey C, and white H. The dashed lines show the calculated anisotropy axes on the dysprosium fragments and the arrows show the ordering of local magnetization axes in the ground state of the complex \cite{Luzon,Chibotaru}. (Lower) The spin structure of the Dy$_3$ triangular cluster and the local easy axes orientation in respect of the laboratory $XYZ$ reference frame.} \end{figure}

The ground state of a Dy$^{3+}$ ion, originating from multiplet $^{6}H_{15/2}$ split in a crystal field, is a Kramers doublet that separated from other states with energy gap $\Delta \sim 200$ cm$^{-1}$ \cite{Luzon,Chibotaru}. In these works the wave functions of the ground doublet are found to be very close to pure states $| M_J = \pm 15/2 \rangle$ relative to the local axes of symmetry of each ion. Wave functions of the first excited doublet correspond to the states $| M_J = \pm 13/2 \rangle$. The local easy axes ($z_i$-axes, $i = 1$, 2, 3) lie on a plane of the dysprosium triangle at an angle of $\varphi$ to the bisectors (see fig.~\ref{dysp}). The most favorable configuration corresponds to $\varphi = 90^{\circ}$ \cite{Chibotaru}.

States of the Dy$^{3+}$ ion system in an external magnetic field under the approximation of the ground doublet are determined by Hamiltonian \cite{Chibotaru}
    \begin{equation} \label{ham}
        {\cal H} = - \sum_{i < k} j_{zz} S_{z_i} S_{z_k} - \mu_B \sum_{i} g_z H_z S_{z_i},
    \end{equation}
where $S = 1/2$ is the effective spin of the ground doublet. The best fit parameters are the effective exchange constant $j_{zz} = 10.6$ K and the effective gyromagnetic factor $g_z = 20.7$ \cite{Luzon}. The eigenfunctions of the Hamiltonian in eq.~(\ref{ham}) are the following vectors:
    \begin{eqnarray*}
        \Phi_{0} = |+++\rangle, \Phi_{1} = |++-\rangle, \\
        \Phi_{2} = |+-+\rangle, \Phi_{3} = |-++\rangle, \\
        \Phi_4 \equiv \bar{\Phi}_{3} = |+--\rangle, \Phi_5 \equiv \bar{\Phi}_{2} = |-+-\rangle, \\
        \Phi_6 \equiv \bar{\Phi}_{1} = |--+\rangle, \Phi_7 \equiv \bar{\Phi}_{0} = |---\rangle,
    \end{eqnarray*}
where ''$+$'' and ''$-$'' are the signs of vector ${\bf J}$ projection of $i$-th ion onto local $z_i$-axis ($i = 1, 2, 3$). Functions $\Phi_4, \ldots, \Phi_7$ are Kramers-conjugate to the $\Phi_3, \ldots, \Phi_0$ relatively.

The spin ordering of the Dy$_3$ nanocluster is characterized in terms of spin chirality \cite{Luzon}. It is clear that the spins in the states $\Phi_i$ ($i=0,1,2,3$) and their conjugates $\Phi_i$ ($i=4,5,6,7$) are inversely twisted, i.e.~the states have the opposite chirality. The natural physical quantity associated with spin chirality in this case is the $t$-odd polar vector of toroidal moment, which corresponds to the first term in toroidal multipole expansion of an arbitrary electric current distribution. It reads as
    $$ {\bf T} = \frac{1}{10c} \int \left( ({\bf j} {\bf r} ) {\bf r} - 2 r^2 {\bf j} \right) d^3 {\bf r}, $$
where ${\bf j} ({\bf r}, t)$ is an electric current density \cite{Dubovik}. In the case of the Dy$_3$ cluster, then we deal with localized magnetic moments, the expression can be transformed to
    $$
    	{\bf T} = \frac{1}{2} g_J \mu_B \sum_{i} [{\bf r}_i \times {\bf J}_i],
    $$
where ${\bf r}_i$ is the radius-vector which connects the center of the triangle with $i$-th apex (see fig.~\ref{dysp}).

\begin{table} \caption{\label{one} Toroidal and magnetic moments}
\begin{ruledtabular} \begin{tabular}{ccc}
  		$\Phi$ & $\tau_z$ & ${\bf m}$ \\
    \hline
    	$\Phi_0$ & $3\sin\varphi$ & $(0;0;0)$ \\
       	$\Phi_1$ & $\sin\varphi$ & $\left( -2\cos\left(\varphi+\frac{\pi}{3}\right); -2\sin\left(\varphi+\frac{\pi}{3}\right); 0 \right)$ \\
       	$\Phi_2$ & $\sin\varphi$ & $\left( -2\cos\left(\varphi-\frac{\pi}{3}\right); -2\sin\left(\varphi-\frac{\pi}{3}\right); 0 \right)$ \\
       	$\Phi_3$ & $\sin\varphi$ & $(2\cos\varphi; 2\sin\varphi; 0)$ \\
\end{tabular} \end{ruledtabular}
\end{table}

The dimensionless values of the toroidal moment $\tau_z = T_Z / T_0$ and the magnetic moment ${\bf m} = {\bf M} / \mu$ (with $\mu = \frac{15}{2} g_J \mu_B$ and $T_0 = \frac{1}{2} \mu r$) in respect of the laboratory $OXYZ$ frame (see fig.~\ref{dysp}) for the states $\Phi_i$ ($i = 0, 1, 2, 3$) are given in the Table~\ref{one}. The values of $\tau_z$ and ${\bf m}$ for the conjugate states $\Phi_i$ ($i = 4, 5, 6, 7$) are different in sign. Energy levels of $\Phi_i$-states are as follows:
    \begin{equation} \label{ens}
        E_0 = E_7 = -\frac{3}{4} j_{zz}, \ E_i = \frac{1}{4} j_{zz} - \mu {\bf m}_i {\bf H}, \ i = 1, \ldots, 6.
    \end{equation} \medskip

\begin{figure} \centering \includegraphics[scale=0.30]{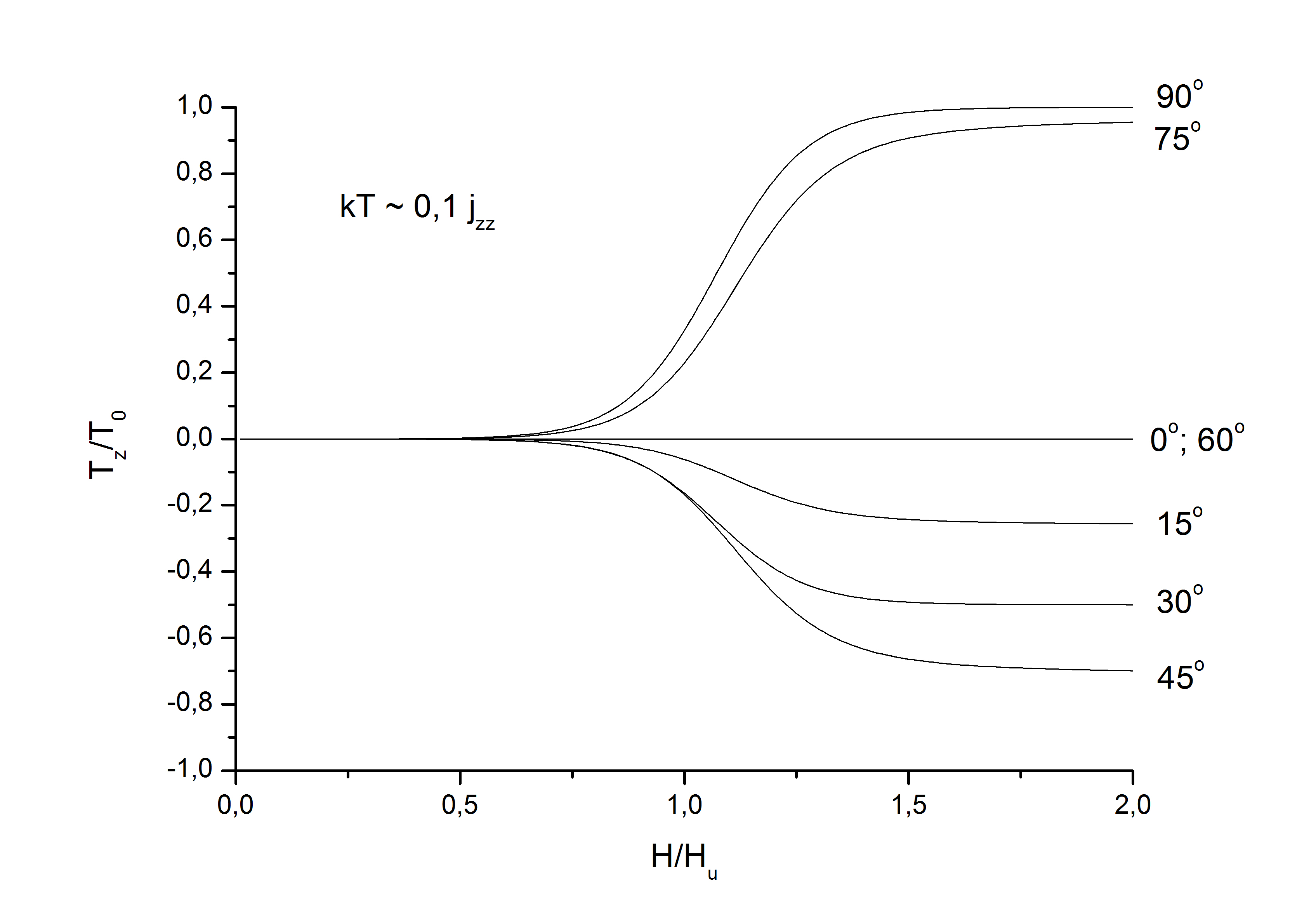} \includegraphics[scale=0.30]{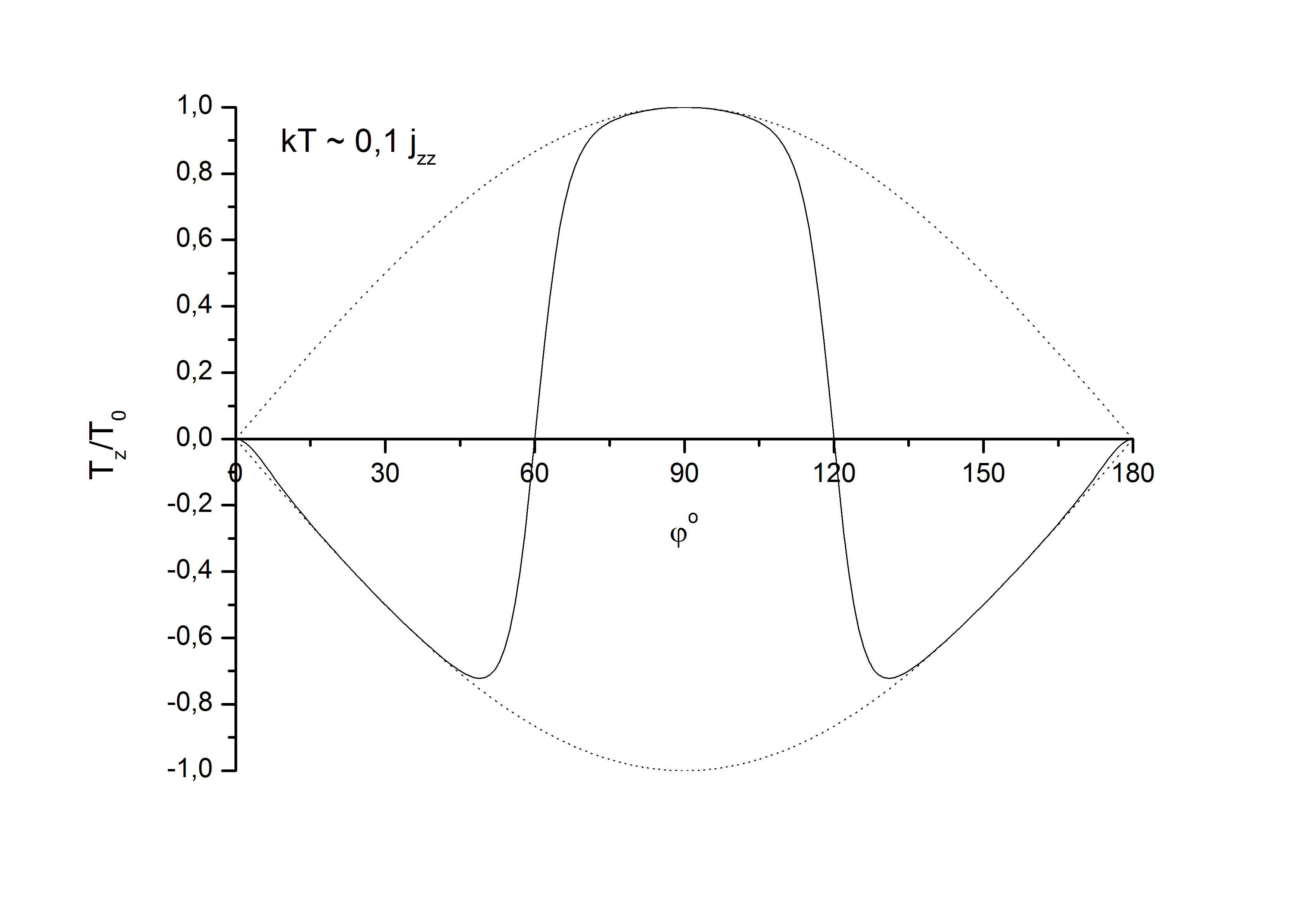} \caption{\label{phi_y} The dependencies of the toroidal moment projection onto the laboratory $OZ$ axis: (upper) on the magnitude of the external magnetic field $H$ for different values of angle $\varphi$; (lower) on the $\varphi$-angle at the strong enough magnetic field $H = 2 H_u$, $H_u = j_{zz} / 2 \mu$. The dotted lines are the envelopes $\pm \sin \varphi$.} \end{figure}

\textbf{3.} In this section we consider in detail the ground state of molecular nanocluster Dy$_3$ as well as the possibility to induce the toroidal moment in it in an external magnetic field at different values of angle $\varphi$, which specifies the orientation of the local magnetization axes in respect of the laboratory $OXYZ$ frame. Let the magnetic field of strength ${\bf H}$ be directed along a side of the dysprosium triangle, i.e. ${\bf H} = (0; H; 0)$. If the field is below the certain threshold value $H_c = H_c(\varphi)$ then the ground state of the system is degenerate state $(\Phi_0,\Phi_7)$. Obviously, the expected value of the toroidal moment in this state equals to zero. On the contrary, in the field strong enough ($H > H_c$) the ground state is not degenerate given the local magnetization axes do not coincide with the triangle sides. In this case, the mean value of the toroidal moment differs from zero, see Table~\ref{two_y}. It is important to underline that this moment is induced by the magnetic field, therefore its sign can be changed by reversing the field direction. The field and angular dependencies of the Dy$_3$ toroidal moment in equilibrium are shown in fig.~\ref{phi_y}.

\begin{table} \caption{\label{two_y} Ground state and toroidal moment of the Dy$_3$ nanocluster in the magnetic field ${\bf H} = (0; H; 0)$}
\begin{ruledtabular} \begin{tabular}{ccccc}
  		$\varphi$ & $H_c(\varphi)$ & $H < H_c$ & $H > H_c$ & $\tau_Z$ \\
    \hline
    	$\varphi = 0^{\circ}$ & $\frac{j_{zz}}{\mu\sqrt{3}}$ & $(\Phi_0,\Phi_7)$ & $(\Phi_2,\Phi_6)$ & $0$ \\
       	$0^{\circ} < \varphi < 60^{\circ}$ & $\frac{j_{zz}}{2\mu\sin\left(\varphi+\frac{\pi}{3}\right)}$ & $(\Phi_0,\Phi_7)$ & $\Phi_6$ & $-\sin\varphi$ \\
       	$\varphi = 60^{\circ}$ & $\frac{j_{zz}}{\mu\sqrt{3}}$ & $(\Phi_0,\Phi_7)$ & $(\Phi_3,\Phi_6)$ & $0$ \\
       	$60^{\circ} < \varphi < 120^{\circ}$ & $\frac{j_{zz}}{2\mu\sin\varphi}$ & $(\Phi_0,\Phi_7)$ & $\Phi_3$ & $\sin\varphi$ \\
       	$\varphi = 120^{\circ}$ & $\frac{j_{zz}}{\mu\sqrt{3}}$ & $(\Phi_0,\Phi_7)$ & $(\Phi_3,\Phi_5)$ & $0$ \\
        $120^{\circ} < \varphi < 180^{\circ}$ & $\frac{j_{zz}}{2\mu\sin\left(\varphi-\frac{\pi}{3}\right)}$ & $(\Phi_0,\Phi_7)$ & $\Phi_5$ & $-\sin\varphi$ \\
       	$\varphi = 180^{\circ}$ & $\frac{j_{zz}}{\mu\sqrt{3}}$ & $(\Phi_0,\Phi_7)$ & $(\Phi_1,\Phi_5)$ & $0$ \\
\end{tabular} \end{ruledtabular}
\end{table}

Thus, depending on the direction of a sufficiently strong external magnetic field it is possible to produce the states of the system with both zero and nonzero  average toroidal moment. \medskip

\textbf{4.} The toroidal moment is of high importance to reveal the spin-electric interactions in the dysprosium cluster and thus to give the description of the quantum MEE in the whole crystal. For this purpose, we consider electric field ${\bf E}$ perturbations and take into account non-Ising corrections in eq.~(\ref{ham}). The electric field perturbation Hamiltonian of a Dy$^{3+}$ ion reads
    \begin{equation} \label{hamv}
        \hat{V} = - \hat{\bf d} {\bf E} + \hat{V}^{odd}_{cr},
    \end{equation}
where $\hat{\bf d} = -e \sum_{i=1}^{N} {\bf r}_i $ is the dipole moment of the ion, $e$ is the elementary charge, $N = 9$ is the number of electrons in $4f$-shell of the Dy$^{3+}$ ion. Crystal field operator $\hat{V}^{odd}_{cr}$ in eq.~(\ref{hamv}) contains only odd harmonics and can be expressed as
    \begin{equation} \label{me:02}
        \hat{V}^{odd}_{cr} = e \sum_{t,\tau,i} a_{t\tau} r^t Y_{t\tau} (\theta_i,\varphi_i),
    \end{equation}
where index $t$ is odd, $Y_{t\tau}(\theta,\varphi)$ are spherical functions, and $a_{t\tau}$ are crystal field parameters.

The simplest form of the crystal field operator $V^{odd}_{cr}$ in eq. \eqref{hamv} corresponds to the first-order term $t = 1$. The environment of each Dy$^{3+}$ ion is of $y \rightarrow -y$ symmetry, thus the resulting inner electric field $E_0 \sim 10^7$ V/cm effective for the ion by the other ions of the cluster is directed along the relevant bisector of the dysprosium triangle. In this case eq. \eqref{hamv} for the $k$-th ion can be expressed as
	\begin{equation} \label{simp}
		V^{odd}_{cr} = e E_0 \sum_{i} \left( x^{(k)}_{i} \sin\varphi + z^{(k)}_{i} \cos\varphi \right).
	\end{equation}

The linear on the strength of the applied electric field corrections to the ion energy levels arise in the second-order perturbation theory with small parameter  $\| V \| / W$, where $\| V \|$ is the norm of the $\hat{V}$-operator and $W$ is the energy difference between ground states and the weight center of excited ion electronic configurations (typically $W \sim 10^5$ cm$^{-1}$ for rare-earth ions).

Making use of the wave function genealogical scheme construction and the quantum theory of angular momentum \cite{Varshalovich} we derive the expression for magnetoelectric operator of the Dy$_3$ complex. The ion electric dipole moment is induced by the magnetic field, which mixes the states of the ground doublet $| M_J = \pm 15/2 \rangle$ with the states of the higher doublet $| M_J = \pm 13/2 \rangle$ (the doublets are separated from each other by the energy $\Delta \sim 200$ см$^{-1}$). Referring the reader to the Appendix for the details of calculations, we give here the final expressions for magnetoelectric corrections to the energy levels of the Dy$_3$ molecular cluster, which are bilinear on $E$ and $H$:
\begin{widetext} \begin{eqnarray} \label{corr}
    \delta E^{me}_{0} & = & 3 A \left[ (H_y E_x - H_x E_y) \cdot \sin\varphi + (H_x E_x + H_y E_y) \cdot \cos\varphi \right], \nonumber \\
    \delta E^{me}_{1,2} & = & \frac{1}{3} \delta E^{me}_{0} +
    2 A \left[ (E_y H_y - E_x H_x) \cdot \cos \left( 3\varphi \pm \frac{\pi}{3} \right) +
    (H_x E_y + H_y E_x) \cdot \sin \left( 3\varphi \pm \frac{\pi}{3} \right) \right], \\
    \delta E^{me}_{3} & = & \delta E^{me}_{0} - \delta E^{me}_{1} - \delta E^{me}_{2}, \,\,\,
    A = - \frac{3}{14} E_0 \frac{(r_{fd}e)^2}{W} \frac{g_J \mu_B}{\Delta}, \nonumber
\end{eqnarray} \end{widetext}
where $r_{fd} = 0.038$ nm is the value of the radial integral for the Dy$^{3+}$ ion \cite{Vedernikov,Judd}. The numerical value of $|A|$ is then estimated to be $3.7 \cdot 10^{-28}$ cm$^3$, therefore given by eq.~\eqref{corr} corrections $\delta E^{me}_i$ reach the value of $\delta E^{me}_i \sim 10^{-3}$ cm$^{-1}$ in external fields $H \sim 1$ T and $E \sim 10^7$ V/cm. The values $\delta E^{me}_{i}$ valid for the conjugate states ($i = 4, 5, 6, 7$) are opposite in sigh with respect of $\delta E^{me}_{i}$ ($i = 0, 1, 2, 3$).

Generally speaking, there can exist another contribution to the cluster polarization through variation in the exchange interaction due to displacement of the ions caused by an external electric field. This electrostrictive mechanism is considered in \cite{Delaney,Bulaevskii} by example of a spin triangle of antiferromagnetically coupled transition-metal (TM) ions. Because the exchange interaction between rare-earth $f$-ions is weaker than that of TM $d$-ions by a factor of hundred, the electrostrictive double-ion mechanism is not effective in rare-earth clusters, while the above-considered single-ion mechanism plays the leading role in the case of rare-earth systems. \medskip


\textbf{5.} In this section we consider an equilibrium MEE in molecular crystal Dy$_3$ in detail. In order to do this, we write an expression for the magnetoelectric contribution to the free energy of the crystal
	$$ {\cal F}_{me} = - k_B T N \ln {\cal Z}, $$
where partition function ${\cal Z}$ is
    $$ {\cal Z} = \sum_{i = 0}^{7} \exp \left( - \frac{E_i + \delta E^{me}_i}{k_B T} \right). $$
The expressions for energies $E_i$ and magnetoelectric corrections $\delta E_i^{me}$ are given by eqs. \eqref{ens} and \eqref{corr} relatively. Quantity $N$ represents the number of Dy$_3$ molecules per volume unit, which is estimated on the basis of Dy$_3$ molecular crystal lattice parameters (see Section 2) to be $8.5 \cdot 10^{20}$~cm$^{-3}$.

An external magnetic field induces the electric polarization in the crystal ($\alpha = x,y,z$)
    \begin{equation} \label{polz}
        P_{\alpha} = - \frac{\partial {\cal F}_{me}}{\partial E_{\alpha}} = -\frac{N \sum_{i = 0}^{7} \frac{\partial (\delta E^{me}_i)}{\partial E_{\alpha}} \cdot e^{-E_i / k_B T}}{\sum_{i = 0}^{7} e^{-E_i / k_B T}},
    \end{equation}
which depends on both magnitude and direction of the external magnetic field vector. \smallskip

\begin{figure} \centering \includegraphics[scale=0.30]{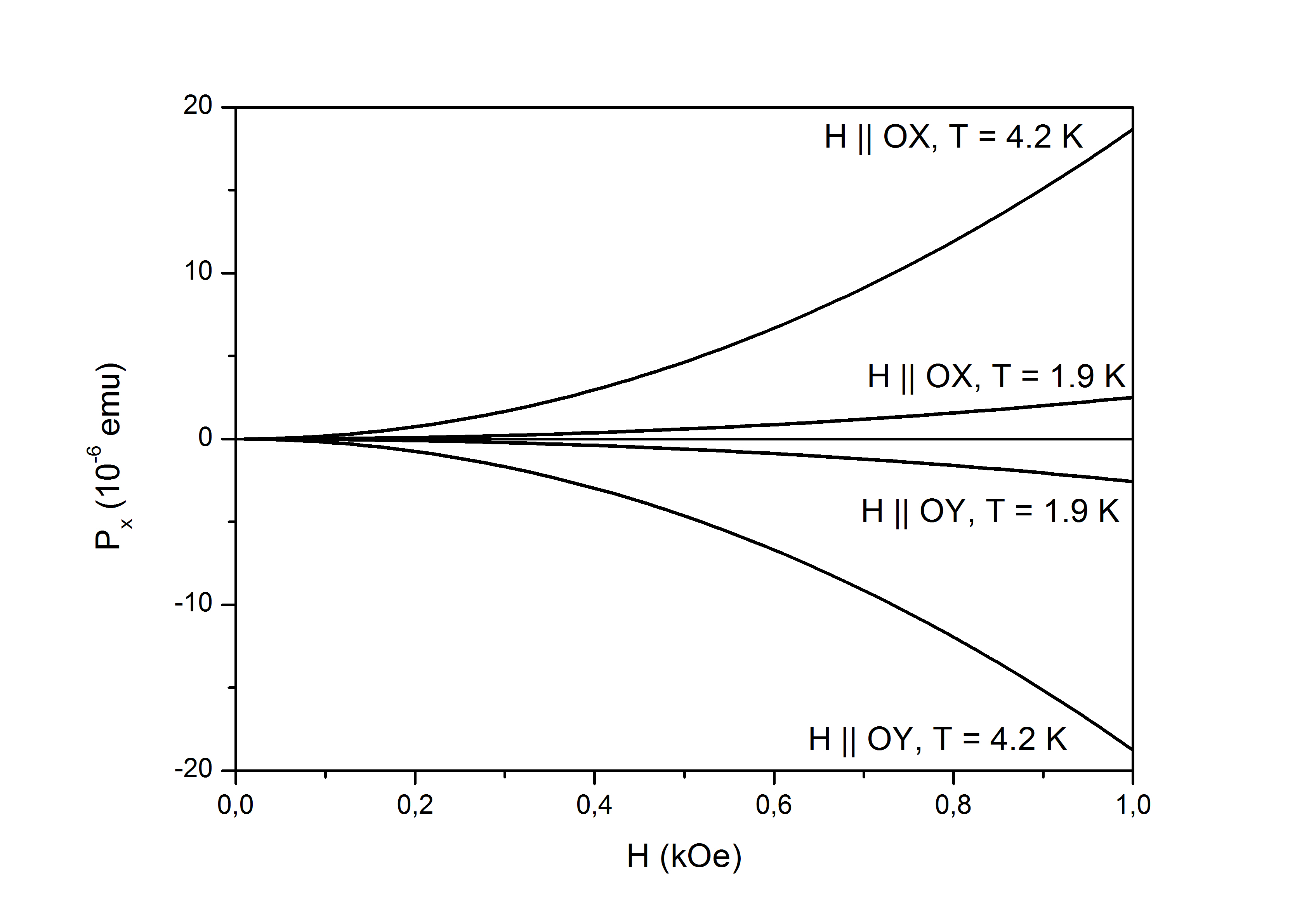} \includegraphics[scale=0.30]{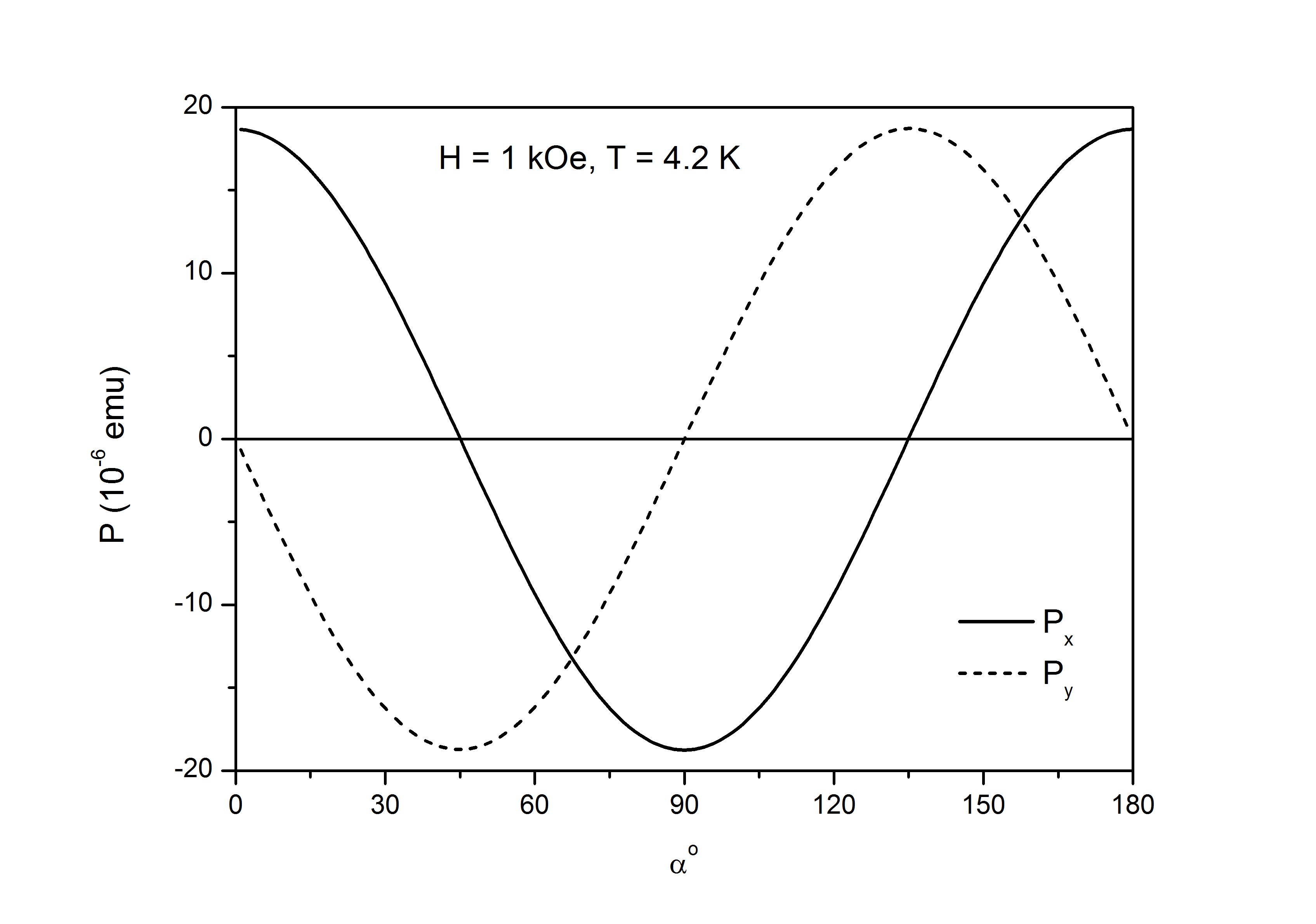} \includegraphics[scale=0.30]{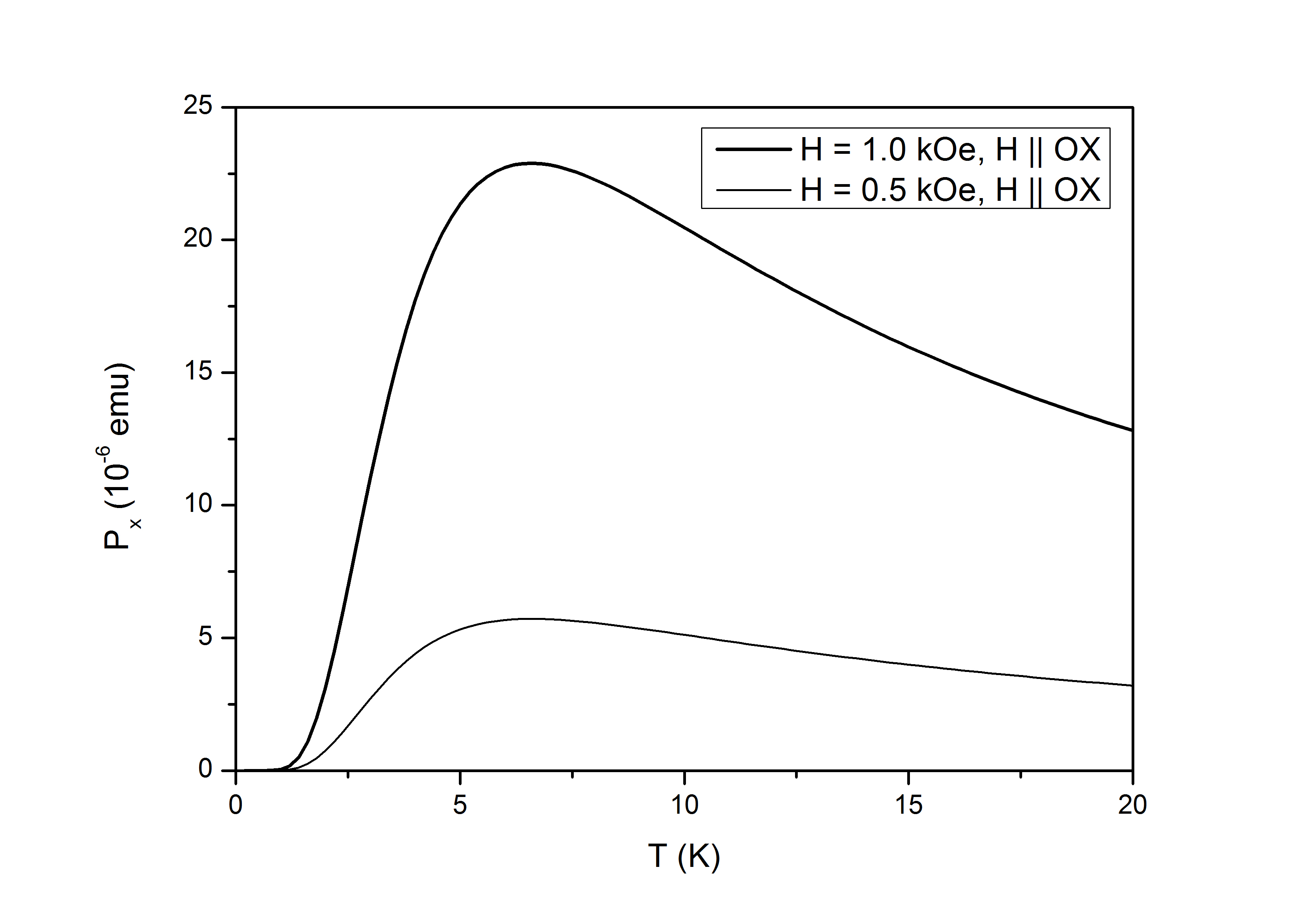} \caption{\label{tiny} (Upper) The isotherms of polarization projection $P_x \sim \pm H^2$ for the case of a weak magnetic field directed along the laboratory $OX$ and $OY$ axes, $P_y = 0$. (Middle) The orientational dependence of polarization projections $P_x \sim \cos 2\alpha$ and $P_y \sim - \sin 2\alpha$ for the case of a weak magnetic field $H = 1$ kOe (at the temperature $T = 4.2$ K), making angle $\alpha$ with the laboratory $OX$-axis. (Lower) The temperature dependence of polarization projection $P_x$ in magnetic fields of $H_x = 1$ kOe and $H_x = 0.5$ kOe for the case of high temperatures $k T >> \mu {\bf m} {\bf H}$.} \end{figure}

\textbf{5.1.} In the limiting case of high temperatures or, equivalently, small fields ($\mu {\bf m} {\bf H} << k T$), for an arbitrary angle $\varphi$, the magnetoelectrical term of the free energy takes a quite simple form
	\begin{equation} \label{stan}
		{\cal F}_{me} = 6 A N \cdot \frac{\mu / k T}{3 + e^{j/kT}} \cdot ( I_1 \cos 2\varphi + I_2 \sin 2\varphi ),
	\end{equation}
where expressions $I_1 = E_x (H_x^2 - H_y^2) - 2 E_y H_x H_y$ and $I_2 = E_y (H_x^2 - H_y^2) + 2 E_x H_x H_y$ stand for the magnetoelectric invariants of point groups $D_3$ and $C_3$ relatively. In the case $\varphi = 90^{\circ}$, eq.~\eqref{stan} contains only standard phenomenological magnetoelectric invariant $I_1$ of a triangle symmetry group ($D_3$). The same invariant determines the thoroughly investigated magnetoelectric properties of rare-earth ferroborates \cite{Zvezdin6,Zvezdin9}. The invariant results in the typical for the paramagnets polarization dependencies on the magnitude ($P \sim H^2$, see Fig. \ref{tiny}a) and the orientation (see Fig. \ref{tiny}b) of the magnetic field strength vector. However, we should point at the unusual for the conventional paramagnets non-monotonic polarization temperature dependence (see Fig. \ref{tiny}c). The main difference is in the direct proportionality between the polarization and the cluster magnetization (not the quadratical as could be expected from the phenomenological analysis).

Obtained in experiment similar curves could be a valuable tool to determine the parameters of the system. Indeed, adjusting the direction of the magnetic field so that one of the polarization components vanishes (say, $P_y \sim \sin(2\alpha - 2\varphi)$), one can easily find the position of the local ion magnetization axes defined by angle $\varphi$. Moreover, knowing the temperature $T_m$ at which the polarization maximum is reached, one can evaluate the exchange constant $j$ from the transcendental equation $$j / k T_m = 1 + 3 e^{-j / k T_m}.$$

\begin{figure} \centering \includegraphics[scale=0.30]{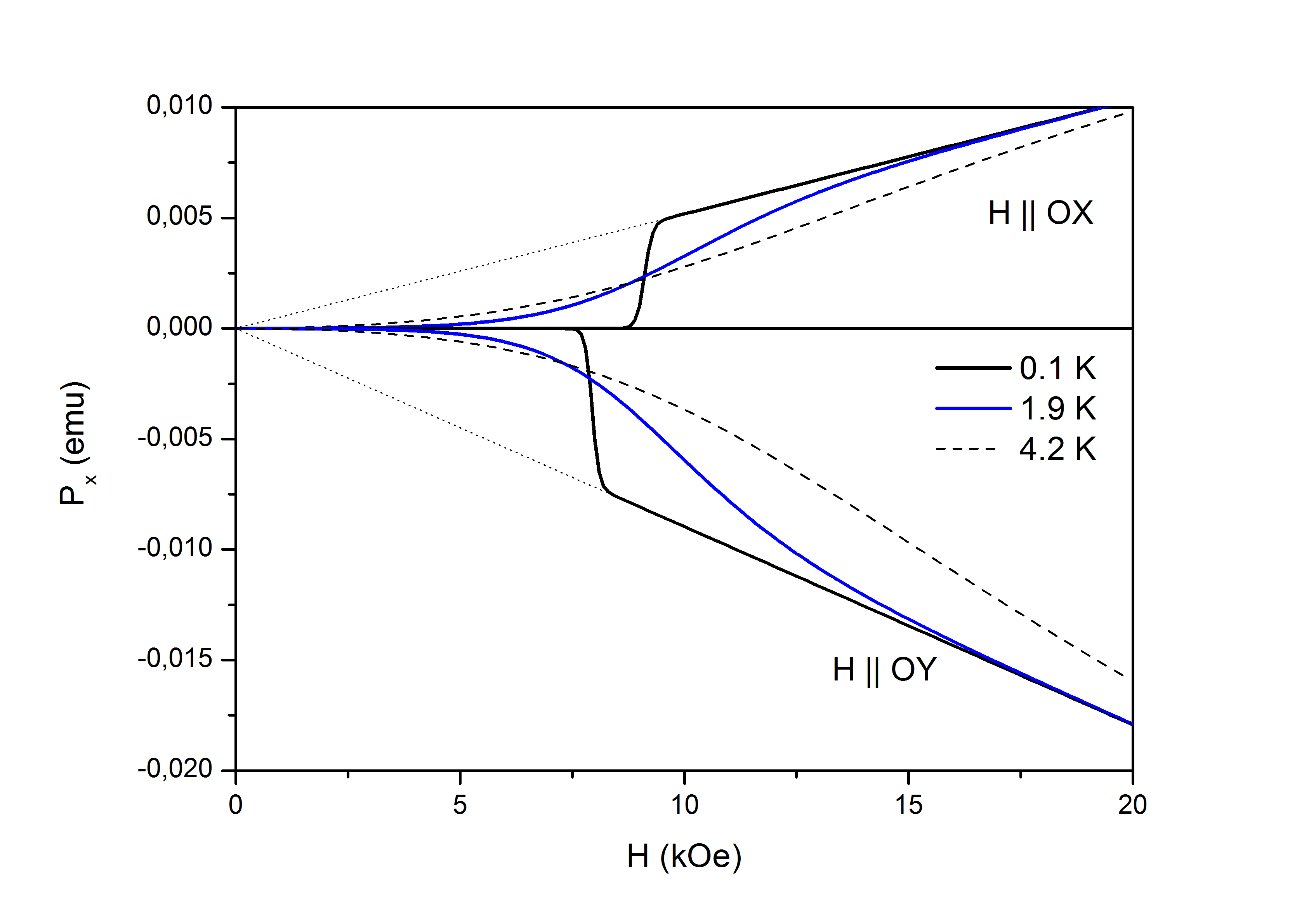} \includegraphics[scale=0.30]{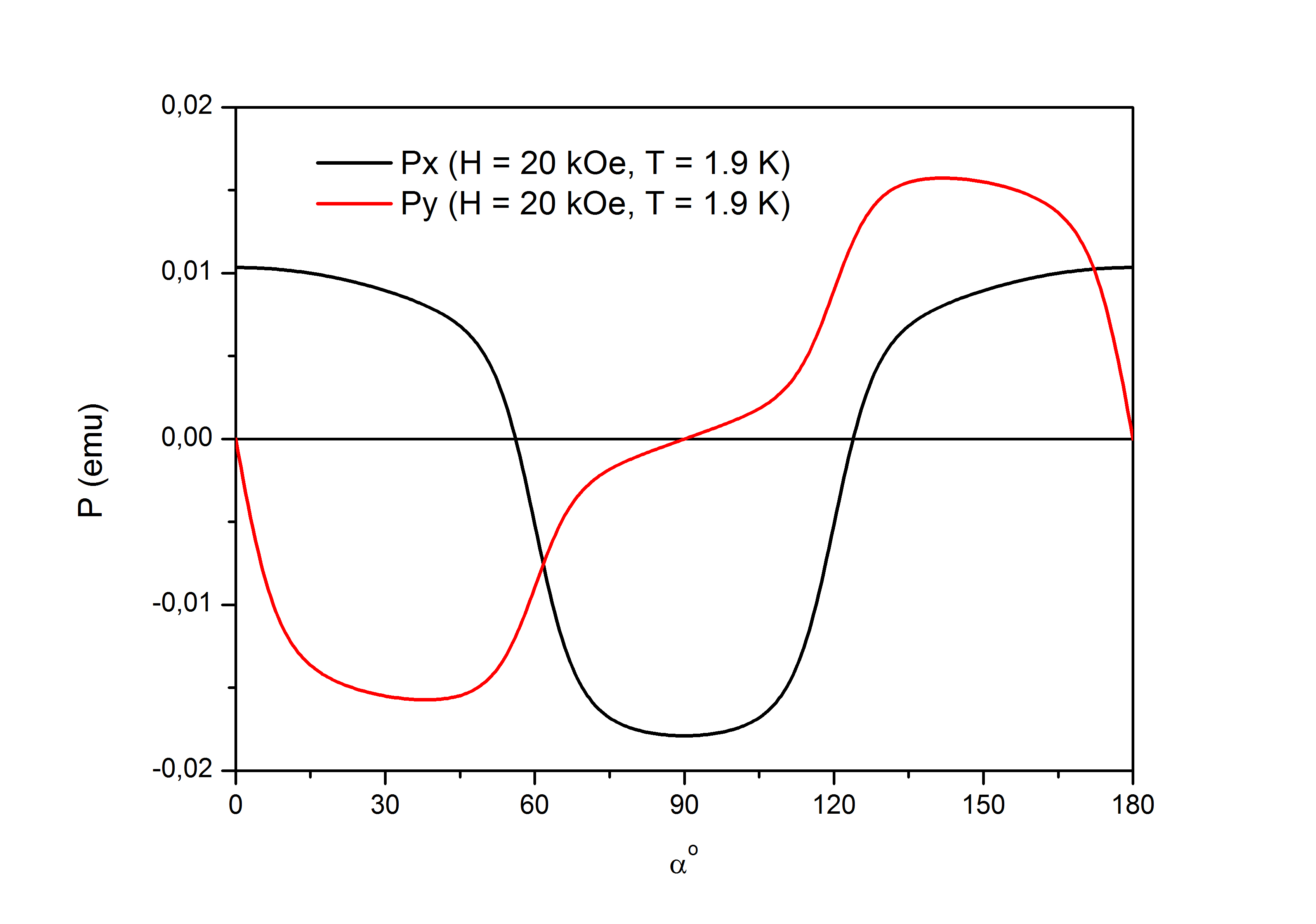} \caption{\label{plot} (Upper) The isotherms of the polarization for the cases of magnetic fields directed along laboratory axes $OX$ (the upper curves) and $OY$ (the lower curves). The dotted lines show the linearity of the MEE in strong enough magnetic fields. (Lower) The orientational dependence of the polarization in the strong magnetic field of $H = 20$ kOe, ${\bf H} = H \left( {\bf e}_x \cos \alpha + {\bf e}_y \sin \alpha \right)$.} \end{figure}

\textbf{5.2.} In the case of strong magnetic fields at low temperatures the polarization behavior is different from that predicted by the phenomenological theory. Indeed, supposed that an external magnetic field is directed along the laboratory $OX$-axis coinciding with a bisector of the dysprosium triangle, the polarization vector is parallel to the field
	\begin{equation} \label{polx}
		P_x = \frac{ 2 \sqrt{3} A N H \sinh(\sqrt{3} \mu H / kT) }{ e^{j_{zz} / k T} + 2 \cosh(\sqrt{3} \mu H / kT) + 1}, \ P_y = 0.
	\end{equation}
If the magnetic field is directed along the laboratory $OY$-axis coinciding with a side of the dysprosium triangle, the components of the polarization changes, so that the polarization vector is perpendicular to the field
	\begin{equation} \label{poly}
		P_x = \frac{ - 3 A N H \sinh( 2 \mu H / k T) }{e^{j_{zz} / kT} + \cosh( 2 \mu H / k T) + 2 \cosh(\mu H / k T)}, \ P_y = 0.
	\end{equation}
Fig.~\ref{plot}a shows the isotherms $P_{\alpha} = P_{\alpha}(H)$, with $\alpha = x, y$, for the mentioned directions of the magnetic field, which reveals the linearity of the MEE at the strong enough fields. The orientational dependencies of polarization are also qualitatively different from conventional ones, see Fig.~\ref{plot}b.

\begin{figure} \includegraphics[scale=0.30]{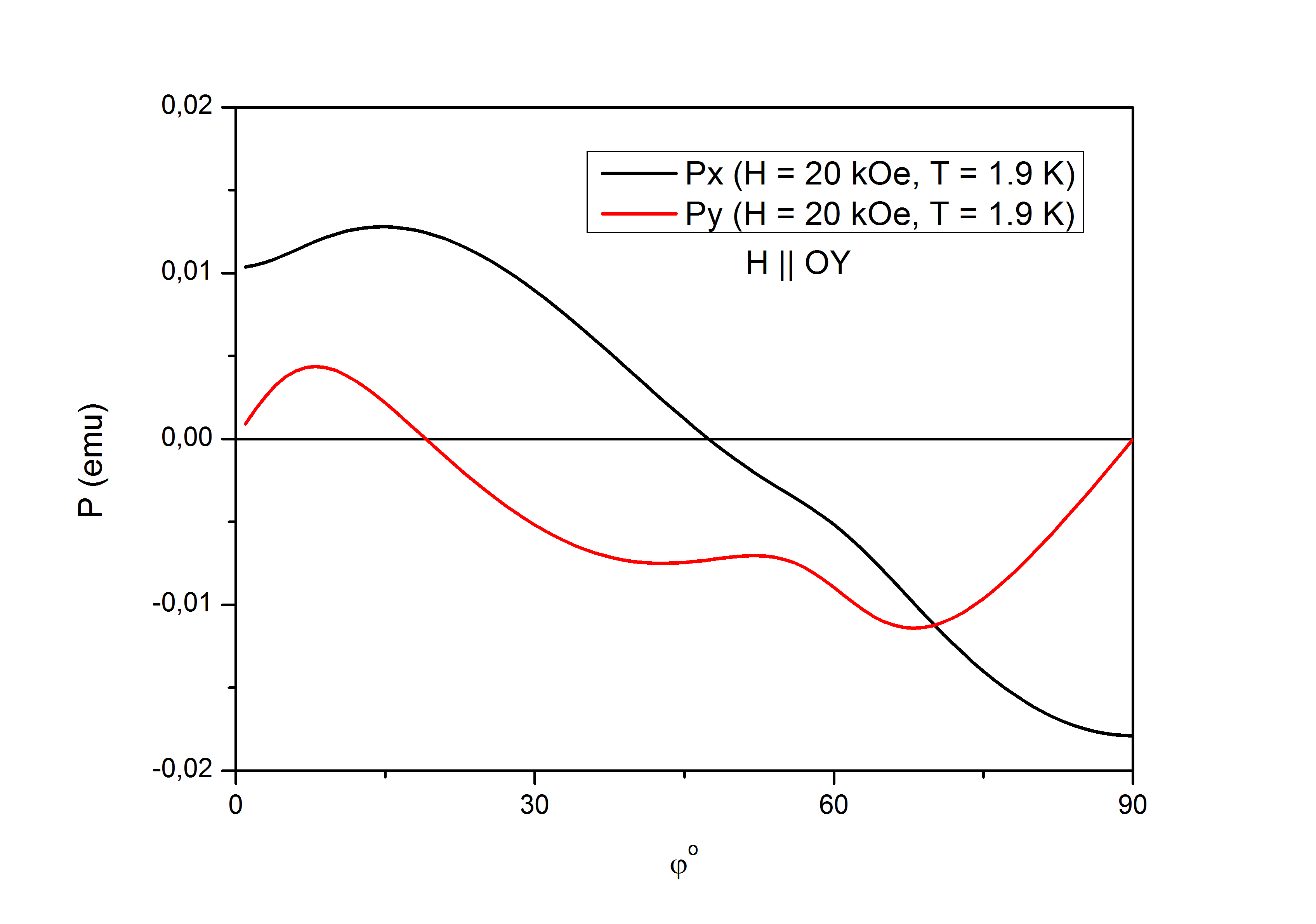} \caption{\label{phis} The polarization dependence on angle $\varphi$, describing the position of the local ion magnetization axes.} \end{figure}

Figs.~\ref{tiny} and \ref{plot} correspond to the most probable value $\varphi = 90^{\circ}$. However, the actual value of angle $\varphi$ is to be specified. This can be done by measuring the polarization components at the fixed magnetic field and temperature. Fig.~\ref{phis} shows the dependence of the components for a range of the $\varphi$ values at the strong magnetic field. \smallskip

\textbf{5.3.} In the weak magnetic field $H < H_{c_1} = j_{zz} / 2 \mu H = 7.9$ kOe regardless of its orientation the ground state of the system is the degenerate state $(\Phi_0, \Phi_7)$. Spin-electric interaction removes the degeneration, thus making the ground state of the Dy$_3$ complex be a doublet with energies [see eq.~\eqref{corr} at $\varphi = 90^{\circ}$]
    $$\delta E = \pm (A / T_0) \left( {\bf T} \cdot [ {\bf E} \times {\bf H} ] \right).$$
To remind, ${\bf T}$ stands here for the toroidal moment, $T_z / T_0 = \tau_z$, see Sec. 2. According to eq.~\eqref{polz}, if the temperature tends to zero then the induced electric polarization is
    \begin{equation} \label{rels} {\bf P} = (A N / T_0) [ {\bf H} \times {\bf T} ], \end{equation}
which is consistent with the expression ${\bf P} \sim [{\bf H} \times {\bf T}]$, derived in \cite{Spaldin,Kopaev} on the base of space-time symmetry considerations.

According to eqs. \eqref{polx} and \eqref{poly}, there is a non-linear MEE in the case of weak magnetic fields but an experimental observation of the effect seems to be difficult due to small interlevel energies $\delta E^{me} \sim 10^{-3}$ cm$^{-1}$. Indeed, the value $\partial P / \partial H$ is less than $10^{-12} - 10^{-11}$ if $H < 5$ kOe and $T \sim 1$ K.

\begin{figure} \centering \includegraphics[scale=0.30]{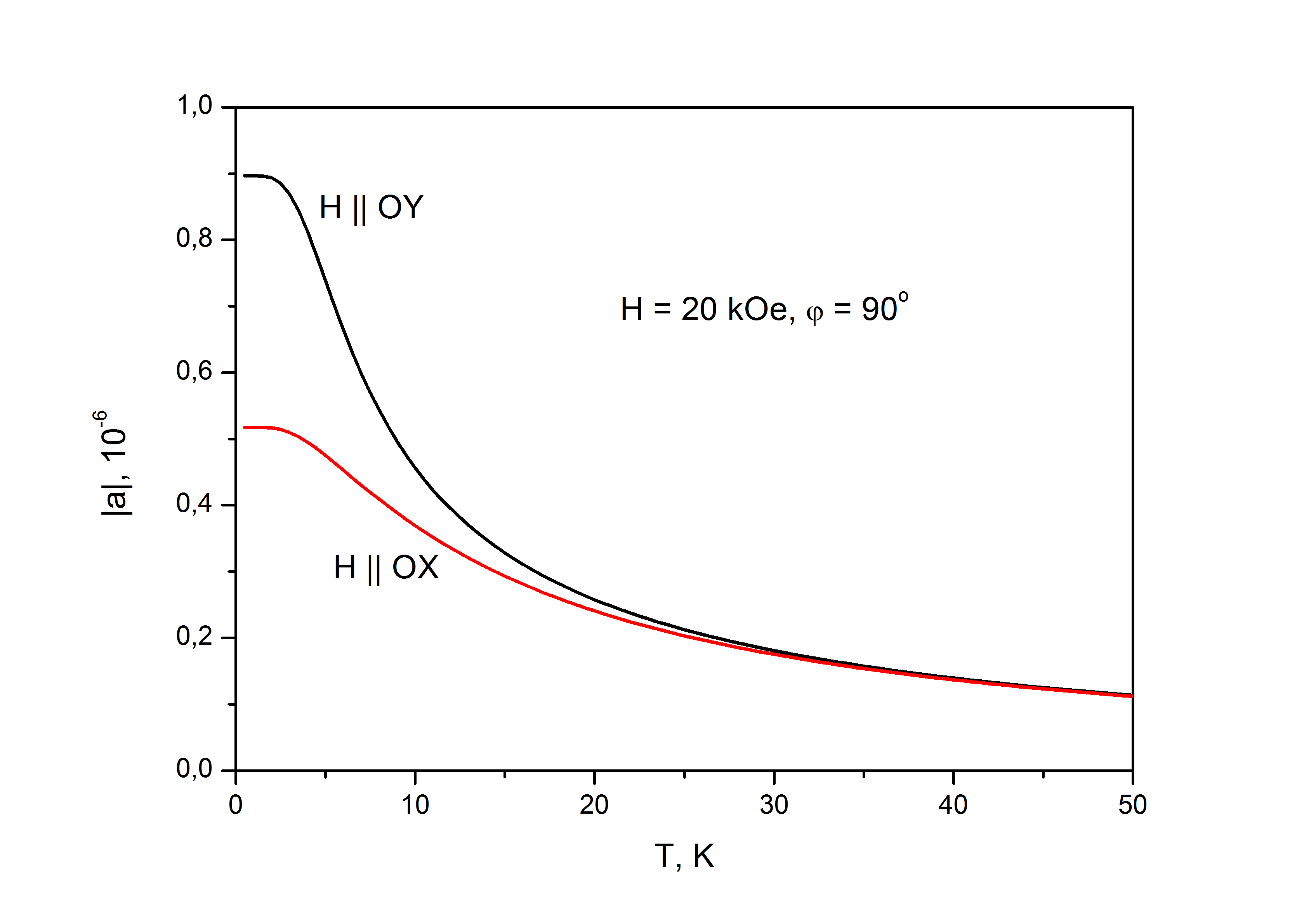} \caption{\label{temp} The temperature dependence of the MEE constant absolute value for the most important orientations of the external magnetic field ($H = 20$ kOe, $\varphi = 90^{\circ}$).} \end{figure}

On the contrary, if the field exceeds the threshold value $H_{c_2} = j_{zz} / \sqrt{3} \mu H = 9.1$ kOe then the ground state is one of the nondegenerate states $\Phi_1 ... \Phi_7$ also independently on the field orientation, except the fields of the levels crossings. In this case, the MEE is observable in experiment. As it follows from Eqs. \eqref{polx} and \eqref{poly}, induced polarization $P = a H$, где $a \sim 10^{-7} - 10^{-6}$ dependent on the external magnetic field orientation. see Fig.~\ref{temp}.

\begin{figure} \centering \includegraphics[scale=0.30]{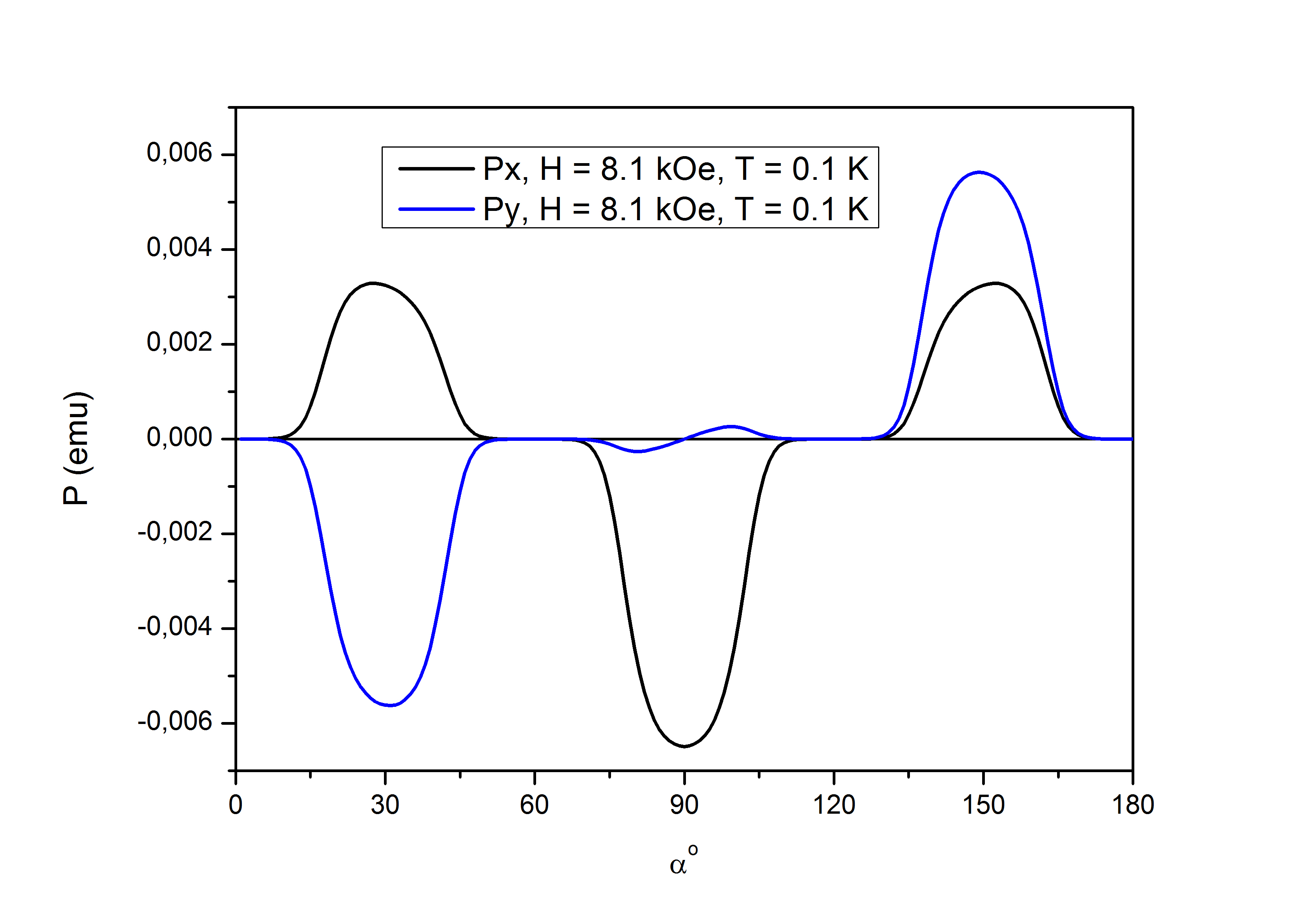} \caption{\label{peak} The plot of the orientational dependencies of the polarization vector projections $P_x$ and $P_y$ in the intermediary magnetic field $H = 8.1$ kOe ($T = 0.1$ K), again ${\bf H} = H \left( {\bf e}_x \cos \alpha + {\bf e}_y \sin \alpha \right)$.} \end{figure}

As for the intermediary region $H_{c_1} < H < H_{c_2}$, the ground state can be both degenerate $(\Phi_0, \Phi_7)$ and nondegenerate $\Phi_n$ $(n = 1 ... 7)$ dependent on the field orientation. This results in the surges of the polarization, see Fig.~\ref{peak}. \medskip

\textbf{6.} The quantum properties of matter at low temperatures are always of special interest. It is remarkable that the peculiar quantum properties of dysprosium molecular clusters can occur in the macroscopic MEE, because they are very weakly coupled, which means that the whole crystal can be treated as an ensemble of particles with zero magnetic moments but nonzero toroidal moments.

First of all, the fundamental possibility of a linear magnetoelectric effect induced by an electric current in the fields $H < H_c$ is worthy of being noted. Indeed, the ground state of a molecule is degenerate unless the magnetic field $H$ does not exceed the threshold value $H_c$. This means zero expected value of the toroidal moment and no linear MEE. The external current $j_z$ removes the degeneration due to interaction with the toroidal moment $(W_{int} = \frac{4\pi}{c} {\bf j} {\bf T})$ producing a state with non-zero toroidal moment and thus resulting in linear MEE.

Not only the mentioned current effect but also quantum dynamical effects are interesting in the context of magnetoelectricity. The slow quantum dynamics of the toroidal moment in SMTs \cite{PRB2011} gives an intriguing possibility to observe non-equilibrium quantum MEE in a crystal of Ising-type rare-earth clusters, when the ground state is the doublet of states with energies (we imply $\varphi = \frac{\pi}{2}$) $E = \pm (3/4) j_{zz} (\Delta/j_{zz})^3$ \cite{PRB2011}, where $\Delta$ is the gap between the energy levels. The wave functions of the states in the doublet are very close to $| \psi_{\pm} \rangle = (| \tau_z = +3 \rangle \pm | \tau_z = -3 \rangle) / \sqrt{2}$. In Dy$^{3+}$ Kramers ions the splitting can be produced with an external magnetic field applied perpendicular to the plane of the ion triangle, $\Delta \sim g_y H$, where $g_y \sim 0.1$ according to the measurements of the magnetization field dependencies at $T = 1.9$ K \cite{Luzon}. Thus, the splitting $\Delta$ is estimated to be 0.05 cm$^{-1}$ in a 1-T magnetic field.

Making use of crossed electric and magnetic fields (or just an electric current) it is possible to prepare the state close to the state characterized by $\tau_z = +3$. In this case, there are oscillations of probability $p(t)$ of finding the system in the state $| \tau_z = -3 \rangle$ after time $t$ \cite{PRB2011}, so that the expected value of the toroidal moment depends on time $\langle \tau_z (t) \rangle = 3 - 6 p(t)$, where $p(t) = \left| \langle \psi_{-} | \psi_{t} \rangle \right|^2$ with $| \psi_t \rangle = \exp \left( - \frac{i}{\hbar} {\cal H} t \right) | \psi_{+} \rangle$.

According to eq.~\eqref{rels}, there exists magnetoelectricity in the toroidal states ${\bf P} = 3 (A N / T_0) [{\bf H} \times {\bf T}]$, therefore the oscillations of $\langle \tau_z(t) \rangle$ result in the oscillations of the electric polarization $P(t) = 3AN (1 - 2p(t)) H$. Note, that the polarization ${\bf P}$ is necessarily perpendicular to the field ${\bf H}$. The frequency of these oscillations is $\nu = \nu_0 (\Delta / j_{zz})^3$ \cite{PRB2011} with $\nu_0 = 48$ GHz and can effectively be controlled for the dysprosium based system by the transversal magnetic field through the value of induced splitting $\Delta$. Given $\Delta \sim 0.05$ cm$^{-1}$, the period of the oscillations is estimated to be 70 $\mu$s. Obtained from the experiment \cite{Luzon} relaxation time $\tau \sim 1$ ms is large enough to observe at least a few oscillations of the polarization. \smallskip


\textbf{7.} In conclusion, the present analysis reveals that chirality of a triangular antiferromagnetic rare-earth nanocluster with Ising-like magnetic anisotropy leads to occurrence of the toroidal moment that allows specific spin-electric interactions and quantum magnetoelectric effect. The resulting quantum structure of the nanocluster is rather rich and versatile for manipulating by external electric and magnetic fields or just by current. Although we have considered mainly the specific rare-earth molecular crystal, based on the Dy$_3$ nanoclusters, many results can be applicable to other Ising-like nanoclusters, e.g. Tb$_3$, Ho$_3$ etc. The rare-earth metal-organic crystals might be the best candidates to observe the macroscopic quantum MEE and other peculiar spin-electric effects owing to the cluster toroidal ground state with long relaxation time. This empowers observation of the quantum linear MEE in a crystal due to the MEE in a single molecule, thus making the old Pierre Curie's idea of MEE in molecules come true. \smallskip

We wish to acknowledge the financial support of the Russian Foundation for Basic Research (projects 10-02-00846, 10-02-01162, 10-02-90475, 11-02-91067, and 11-02-92006). \smallskip


\begin{widetext} \section*{Appendix}

The purpose of this section is the detailed and clear derivation of the spin-electric Hamiltonian of the Dy$_3$ triangular complex and the calculation of the magnetoelectric corrections \eqref{corr} to the energy levels of the system. As mentioned in Sec.~4, the analysis is based on the perturbation theory. The second-order corrections read
    \begin{equation} \label{korr}
        E^{(2)}_g = \sum_{l',e_{l'}} \frac{1}{W_{l'}} \left( \langle g | {\bf d} {\bf E} | e_{l'} \rangle \langle e_{l'} | V^{odd}_{cr} | g \rangle  + \langle g | V^{odd}_{cr} | e_{l'} \rangle \langle e_{l'} | {\bf d} {\bf E} | g \rangle \right).
	\end{equation}
Here $| g \rangle$ are the dysprosium ion states of the ground $l^N$ configuration in a magnetic field ($l = 3$ is the orbital quantum number of rare-earth ions and $N = 9$ is the number of electrons in $4f$-shell of the Dy$^{3+}$ ion), $| e_{l'} \rangle$ are the states from the excited $l^{N-1}l'$ configuration with $l' = l \pm 1$, and $W_{l'}$ is the energy difference between $| e_{l'} \rangle$ and $| g \rangle$ states. The splitting of the $l^{N-1}l'$ configuration levels is neglected. The configuration with $l' = l - 1 = 2$ ($W_{l-1} \sim 10^5$ cm$^{-1}$) is closer to the ground levels than the configuration with $l' = l + 1 = 4$.

The interaction operator of the ion dipole moment with an external electric field in Eq.~\eqref{hamv} can be written in terms of irreducible tensor operators $d_{1\mu}$ ($\mu = 0, \pm1$)
	\begin{equation}
		{\bf d} {\bf E} = \sum_{\mu} (-1)^{\mu} E_{-\mu} d_{1\mu} , \nonumber
	\end{equation}
where $d_{1,\pm1} = \mp (d_x \pm id_y) / \sqrt{2}$, $d_{10} = d_z$, $E_{\pm1} = \mp (E_x \pm iE_y) / \sqrt{2}$, and $E_0 = E_z$.

For the wave functions $| g \rangle$ and $| e_{l'} \rangle$ construction we use the genealogical scheme, in particular $| g \rangle = \sum a_M | J M \rangle$, where $J$ and $M$ are the quantum numbers of the ground multiplet of the ion, $a_M$ are the numerical coefficients given by the following equations
	\begin{equation}
		| J M \rangle = \sum_{M_L M_S} C^{J M}_{L M_L S M_S} | l^N S L M_S M_L \rangle, \nonumber
	\end{equation}
	\begin{equation}
		| l^N S L M_S M_L \rangle = \sum_{S_1 L_1 i} \sum_{M_{S_1} M_{L_1} \mu}
        G^{S L}_{S_1 L_1} \cdot C^{L M_L}_{L_1 M_{L_1} l m} \cdot C^{S M_S}_{S_1 M_{S_1} \frac{1}{2} \mu}
        \cdot \Psi_{S_1 L_1 M_{S_1} M_{L_1}} \cdot (-1)^{N-i} \cdot \psi_{l m \frac{1}{2} \mu} (\xi_i), \nonumber
	\end{equation}
	\begin{equation}
		| e_{l'} \rangle = \frac{1}{\sqrt{N}} \sum (-1)^{N-i} \cdot \Psi_{S_1 L_1 M_{S_1} M_{L_1}} \cdot \psi_{l' m' \frac{1}{2} \mu'} (\xi_i), \nonumber
	\end{equation}
where $C^{R r}_{\nu \mu t \tau}$ and  $G^{S L}_{S_1 L_1}$ stand for the Clebsch-Gordan and genealogical coefficients respectively, and $L_1$, $S_1$, $M_{L_1}$, $M_{S_1}$ are the quantum numbers of the initial therm.

The known \cite{Varshalovich} identity $\sum_{nm} C_{1 \mu t \tau}^{n m} C_{1 \nu t q}^{n m} = \delta_{\mu \nu} \cdot \delta_{\tau q}$ yields
	\begin{equation} \label{ang1}
		\sum_{e_{l'}} \langle g | d_{1\mu} | e_{l'} \rangle \langle e_{l'} | r^t Y_{t\tau} | g \rangle =
        \sum_{n m} C^{n m}_{1 \mu t \tau} \langle g | \left( d_1 \otimes \left( r^t Y_t \right) \right)_{n m} | g \rangle,
	\end{equation}
where
	\begin{equation} \label{pent}
		\langle g | (d_1 \otimes (r^t Y_t))_{n m} | g \rangle = \sum_{e_{l'} \nu q} C^{n m}_{1 \nu t q}
		\langle g | d_{1 \nu} | e_{l'} \rangle \langle e_{l'} | r^t Y_{t q} | g \rangle.
	\end{equation}
The matrix elements $\langle g | T_{kq} | e_{l'} \rangle$ of irreducible tensor operator $T_{kq} = \sum_i t_{kq} (i)$ read
	\begin{equation} \label{bulk}
		\langle g | T_{kq} | e_{l'} \rangle = \sqrt{N} \sum_{M, M_L, M_S, S_1, L_1, M_{S_1}, M_{L_1}, m_l, m_{l'}, \mu}
		a^{*}_M \cdot C^{JM}_{L M_L S M_S} \cdot G^{S L}_{S_1 L_1} \cdot C^{L M_L}_{L_1 M_{L_1} l m_l} \cdot C^{S M_S}_{S_1 M_{S_1} 1/2 \mu}
		\left\langle l m_l \left| t_{kq} \right| l' m_{l'} \right\rangle,
	\end{equation}
where
	$\left\langle l m_l \left| t_{kq} \right| l' m_{l'} \right\rangle = C^{l m_l}_{l' m_{l'} k q}
	\frac{\langle l || t_{kq} || l' \rangle}{\sqrt{2l + 1}}$, $\langle l || t_{kq} || l' \rangle$ is the reduced matrix element.
	
Substituting eq.~\eqref{bulk} into Eq~\eqref{pent} we obtain
	\begin{eqnarray} \label{zudk}
		\left\langle g \left| \left\{ d_1 \otimes ( r^t Y_t ) \right\}_{nm} \right| g \right\rangle =
		-eN \sum C^{n m}_{1 \nu t q} \cdot C^{l m_l}_{l' m_{l'} 1 \nu} \cdot C^{l' m_{l'}}_{l m'_l t q} \cdot a^{*}_M \cdot a_{M'} \times \nonumber \\
		\times C^{J M}_{L M_L S M_S} \cdot C^{J M'}_{L M'_L S M_S} \cdot C^{L M_L}_{L_1 M_{L_1} l m_l} \cdot	C^{L M'_L}_{L_1 M'_{L_1} l m'_l} \cdot
		\left( G^{S L}_{S_1 L_1} \right)^2 \cdot \frac{ \langle l || r_1 || l' \rangle \langle l' || r^t Y_t || l \rangle }{\sqrt{(2l + 1)(2l' + 1)}}.
	\end{eqnarray}
According to \cite{Varshalovich}, the reduced matrix elements in Eq.~\eqref{zudk} can be written as $\langle l || r_1 || l' \rangle = r_{ll'} \sqrt{2l' + 1} C^{l0}_{l'010}$ and $\langle l' || r^t Y_t || l \rangle = (r^t)_{ll'} \sqrt{\frac{(2t + 1)(2l + 1)}{4\pi}} C^{l'0}_{l0t0}$, where $(r^t)_{ll'} = \langle l | r^t | l' \rangle$ are the radial integrals.

Summing up of the first three factors in Eq.~\eqref{zudk} over $\nu$, $q$, and $m_{l'}$ results in (see \cite{Varshalovich})
	\begin{equation} \label{trys}
		\sum_{\nu q m_{l'}} C^{n m}_{1 \nu t q} \cdot C^{l m_l}_{l' m_{l'} 1 \nu} \cdot C^{l' m_{l'}}_{l m'_l t q} =
        (-1)^n \sqrt{(2n+1)(2l'+1)} C^{l m_l}_{l m'_l n m} \left\{ \begin{array}{*{20}{c}} t & 1 & n \\ l & l & l' \end{array} \right\},
	\end{equation}
where $\left\{ \begin{matrix} t & 1 & n \\ l & l & l' \\ \end{matrix} \right\}$ are the $6j$-symbols with even indexes $n$.

From the equation $\langle l m_l | Y_{nm} | l m'_l \rangle  = C^{l m_l}_{l m'_l n m} \sqrt{\frac{2n + 1}{4\pi}} C^{l 0}_{l 0 R 0}$ we get $C^{l m_l}_{l m'_l n m}$ and then from eqs.~\eqref{trys} and \eqref{zudk} we find that
	\begin{equation}
		\langle g | (d_1 \otimes ( r_t Y_t))_{nm} | g \rangle =
        A(ll',tn) \left\langle g \left| \sum_{i=1}^{N} Y_{nm} (\theta_i; \varphi_i) \right| g \right\rangle
		A(ll',tn) \alpha_n \langle g | Y_{nm} ({\bf J}) | g \rangle, \nonumber
	\end{equation}
	\begin{equation}
		A(ll',tn) = -e r_{ll'} (r^t)_{ll'} \frac{C^{l 0}_{l' 0 1 0 } \cdot C^{l' 0}_{l 0 t 0}}{C^{l 0}_{l 0 n 0}} \nonumber \\
		\left\{ \begin{array}{*{20}{c}} t & 1 & n \\ l & l & l' \end{array} \right\}
		\sqrt{(2l' + 1)(2l + 1)}, \nonumber
	\end{equation}
where $\alpha_n$ are the Stevence parameters.

After substituting eq.~\eqref{ang1} into Eq.~\eqref{korr}, we arrive to the contribution of an ion to the MEE operator projected onto the space of the ground multiplet functions
	\begin{equation} \label{hmee}
		{\cal H}_{me} = 2 \Re \sum_{\mu t \tau} \sum_{n m} (-1)^{\mu} \cdot E_{-\mu} \cdot a_{t \tau} \cdot
        A(ll', tn) \cdot W^{-1}_{l'} \cdot C^{n m}_{1 \mu t \tau} \cdot \alpha_n \cdot \hat{Y}_{n m} (J).
	\end{equation}
The effective magnetoelectric operator of the dysprosium complex can be obtained if one sums up the contributions given by eq.~\eqref{hmee} over all three dysprosium ions,
	\begin{equation} \label{meeq}
		{\cal H}_{me} = \sum_{\alpha n m k} B^{\alpha}_{nm} E^{(k)}_{\alpha} Y^{(k)}_{nm} (J),
	\end{equation}
where the index $k = 1, 2, 3$ refers to Dy$^{3+}$ ions, $\beta = x, y, z$,	
	\begin{eqnarray}
		B^{x}_{nm} = \sqrt{2} \sum_{t \tau l'} \Re a_{t\tau} \cdot \left( C^{nm}_{1 -1 t \tau} - C^{n m}_{1 +1 t \tau} \right)
		\cdot A(ll',tn) \cdot W^{-1}_{l'} \cdot \alpha_n, \nonumber \\
		B^{y}_{nm} = -\sqrt{2} \sum_{t \tau l'} \Im a_{t\tau} \cdot \left( C^{nm}_{1 -1 t \tau} + C^{n m}_{1 +1 t \tau} \right)
		\cdot A(ll',tn) \cdot W^{-1}_{l'} \cdot \alpha_n, \nonumber \\
		B^{z}_{nm} = 2 \sum_{t \tau l'} \Re a_{t\tau} \cdot C^{n m}_{1 0 t \tau} \cdot A(ll',tn) \cdot W^{-1}_{l'} \cdot \alpha_n. \nonumber
	\end{eqnarray}

In order to avoid analysis of excessively bulky expressions we consider the simplest form of the crystal field operator $V_{cr}^{odd}$, see eq.~\eqref{simp}, and write the coordinates of the local axes explicitly
	$$
		\begin{matrix}
			{\bf e}_{x_1} = (\sin\varphi, -\cos\varphi, 0), & {\bf e}_{y_1} = (0, 0, -1), & {\bf e}_{z_1} = (\cos\varphi, \sin\varphi, 0), \\
			{\bf e}_{x_2} = \left( \cos \left( \varphi + \frac{\pi}{6} \right), \sin \left( \varphi + \frac{\pi}{6} \right), 0\right), &
			{\bf e}_{y_2} = (0, 0, -1), &
			{\bf e}_{z_2} = \left( -\sin \left( \varphi + \frac{\pi}{6} \right), \cos \left( \varphi + \frac{\pi}{6} \right), 0\right), \\
			{\bf e}_{x_3} = \left( -\cos \left( \varphi - \frac{\pi}{6} \right), \sin \left( \varphi - \frac{\pi}{6} \right), 0\right), &
			{\bf e}_{y_3} = (0, 0, -1), &
			{\bf e}_{z_3} = \left( \sin \left( \varphi - \frac{\pi}{6} \right), -\cos \left( \varphi - \frac{\pi}{6} \right), 0\right).
		\end{matrix}
	$$	
From eqs. \eqref{hmee} and \eqref{simp} after some calculations one obtains that
  \begin{equation} \label{me:10}
  	\hat{\cal H}_{me} = \frac{E_0 \alpha_2 C}{2l+1} \sum_{k \alpha} E^{(k)}_{\alpha}
  	\left( Q^{(k)}_{x\alpha}(J) \sin\varphi + Q^{(k)}_{z\alpha}(J) \cos\varphi \right),
  \end{equation}
where $Q_{\alpha\beta}(J) = \frac{1}{2} \left( J_{\alpha} J_{\beta} + J_{\beta} J_{\alpha} - \frac{2}{3} \delta_{\alpha \beta} J(J+1) \right)$ are the quadrupole moment components of the Dy$^{3+}$ ion $f$-shell, coefficient $\alpha_2 = - \frac{2}{5 \cdot 7 \cdot 9}$ is the Stevence parameter, and $C = (2l+3) (r_{fd}e)^2 / W$. 		

As it was already mentioned, the ground state of a dysprosium ion in the cluster is a doublet that is very close to $|\pm15/2\rangle$ (relative to the local symmetry axes). Excited doublet $|\pm13/2\rangle$ is well separated from the ground one by the gap of $\Delta \sim 200$ cm$^{-1}$. External magnetic field $H$ entangles states $|\pm13/2\rangle$ with states $|\pm15/2\rangle$, thus making the ground ion states in the field to be as
    $$   \left| g^{(k)}_{\pm} \right\rangle = \left| \pm \frac{15}{2} \right\rangle -
        \frac{\mu (H^{(k)}_{x} \pm i H^{(k)}_{y})}{\Delta \sqrt{15}} \left| \pm \frac{13}{2} \right\rangle.
    $$
In this case, the only nonzero linearly depending on magnetic field matrix element of quadrupole moment is the component $\hat{Q}_{xz}(J)$
    \begin{equation} \label{me:12}
        \left\langle g^{(k)}_{\pm} \right| \hat{Q}^{(k)}_{xz}(J) \left| g^{(k)}_{\pm} \right\rangle =
        \mp \frac{7 \mu H^{(k)}_{x}}{\Delta}.
    \end{equation}

Substituting eq.~\eqref{me:12} for $\hat{Q}_{xz}(J)$ in eq.~\eqref{me:10}, one obtains bilinear on $E$ and $H$ corrections to the energy levels $E_i$ of the Dy$_3$ molecular complex:
    \begin{equation} \label{me:13}
        \delta E^{me}_{i} = - \frac{E_0 \alpha_2 C}{2l+1} \cdot \frac{7 \mu}{\Delta} \cdot
        \sum_{k} \sigma^{(k)}_{z} (i) H^{(k)}_{x} \left( E^{(k)}_{z} \sin\varphi + E^{(k)}_{x} \cos\varphi \right),
    \end{equation}
where $i = 0, \ldots, 7$. Symbol $\sigma^{(k)}_{z} (i)$ has been introduced for the sign of a magnetic quantum number of $k$-th Dy$^{3+}$ ion in the $\Phi_i$ state $(\sigma^{(k)}_{z} = \pm 1)$. After summation over $k$-index in eq.~\eqref{me:13}, we finally arrive to eq.~\eqref{corr}.

\end{widetext}



\begin{thebibliography}{99}

\bibitem{Hasan} M. Z. Hasan and C. L. Kane, Rev. Mod. Phys. \textbf{82}, 3045 (2010)

\bibitem{Castelnovo} C. Castelnovo, R. Moessner, and S. L. Sondhi, Nature \textbf{451}, 42 (2008)

\bibitem{Schmid} H. Schmid, J. Phys.: Condens. Matter, \textbf{20}, 434201 (2008)

\bibitem{Spaldin} N. A. Spaldin, M. Feibig, and M. Mostovoy, J. Phys.: Cond. Mat. \textbf{20} 434203 (2008)

\bibitem{Kopaev} Yu. V. Kopaev, Phys. Usp. \textbf{52}, 1111 (2009)

\bibitem{Dubovik} V. M. Dubovik and V. V. Tugushev, Phys. Rep. \textbf{187}, 145 (1990)

\bibitem{Popov} Yu. F. Popov, A. M. Kadomtseva, G. P. Vorobiev at al., J. Exp. Theor. Phys. \textbf{87}, 146 (1998)

\bibitem{Sannikov} D. G. Sannikov, Ferroelectrics \textbf{219}, 177 (1998)

\bibitem{Krotov} Yu. F. Popov, A. M. Kadomtseva, D. V. Belov at al., JETP Lett. \textbf{69}, 330 (1999)

\bibitem{Lebeugle} D. Lebeugle, A. Mougin, M. Viret et al., Phys. Rev. Lett. \textbf{103}, 257601 (2009)

\bibitem{Kleemann} W. Kleemann, Physics \textbf{2}, 105 (2009)

\bibitem{Chen} Xi Chen, A. Hochstrat, P. Borisov, and W. Kleemann, Appl. Phys. Lett. \textbf{89}, 202508 (2006)

\bibitem{Tang} J. Tang, I. Hewitt, N. T. Madhu et al., Angew. Chem. Int. Ed. \textbf{45} 1729 (2006)

\bibitem{Soncini} A. Soncini and L. F. Chibotaru, Phys. Rev. B, \textbf{77}, 220406 (2008)

\bibitem{EPL2009} A. I. Popov, D. I. Plokhov, and A. K. Zvezdin, Europhys. Lett. \textbf{87}, 67004 (2009)

\bibitem{Trif} M. Trif, F. Troiani, D. Stepanenko, and D. Loss, Phys. Rev. B, \textbf{82}, 045429 (2010)

\bibitem{Chibotaru} L. F. Chibotaru, L. Ungur, and A. Soncini, Angew. Chem. Int. Ed. \textbf{47} 4126 (2008)

\bibitem{Luzon} J. Luzon, K. Bernot, I. J. Hewitt et al., Phys. Rev. Lett. \textbf{100}, 247205 (2008)

\bibitem{EPAPS} EPAPS Document No. E-PRLTAO-100-067825, see http://www.aip.org/pubservs/epaps.html

\bibitem{Varshalovich} D. Varshalovich, A. Moskalev, V. Khersonskii, \textit{Quantum theory of angular momentum}, World Scientific (1989)

\bibitem{Vedernikov} N. F. Vedernikov, A. K. Zvezdin, R. Z. Levitin, and A. I. Popov, Sov. Phys. JETP, \textbf{66}, 1233 (1987)

\bibitem{Judd} B. R. Judd, Phys. Rev. \textbf{127}, 750 (1962)

\bibitem{Delaney} K. T. Delaney, M. Mostovoy, and N. A. Spaldin, Phys. Rev. Lett. \textbf{102} 157203 (2009)

\bibitem{Bulaevskii} L. N. Bulaevskii, C. D. Batista, M. V. Mostovoy, and D. I. Khomskii, Phys. Rev. B \textbf{78} 024402 (2008)

\bibitem{Zvezdin6} A. K. Zvezdin, G. P. Vorob'ev, A. M. Kadomtseva et al., J. Exper. Theor. Phys. Lett. \textbf{83}, 509 (2006)

\bibitem{Zvezdin9} A. K. Zvezdin, A. M. Kadomtseva, Yu. F. Popov et al., J. Exper. Theor. Phys. \textbf{109}, 68 (2009)

\bibitem{PRB2011} D.I. Plokhov, A.K. Zvezdin, and A.I. Popov, Phys. Rev. B \textbf{83}, 184415 (2011)

\end{thebibliography}
\end{document}